\documentclass[aps,prl,twocolumn,superscriptaddress,groupedaddress]{revtex4}  
\usepackage{graphicx}  
\usepackage{dcolumn}   
\usepackage{bm}        
\usepackage{amssymb}   
\usepackage{sidecap}   
\usepackage[dvips]{color}	
\begin{document}

\widetext
\title{Soliton Instabilities and Vortex Streets Formation in a Polariton Quantum Fluid}

\affiliation{ICMP, Ecole Polytechnique F\'ed\'erale de Lausanne (EPFL), 1015 Lausanne, Switzerland}

\author{G.~Grosso} \affiliation{ICMP, Ecole Polytechnique F\'ed\'erale de Lausanne (EPFL), 1015 Lausanne, Switzerland}
\author{G.~Nardin} \affiliation{ICMP, Ecole Polytechnique F\'ed\'erale de Lausanne (EPFL), 1015 Lausanne, Switzerland}
\author{F.~Morier-Genoud} \affiliation{ICMP, Ecole Polytechnique F\'ed\'erale de Lausanne (EPFL), 1015 Lausanne, Switzerland}
\author{Y.~L\'eger} \affiliation{ICMP, Ecole Polytechnique F\'ed\'erale de Lausanne (EPFL), 1015 Lausanne, Switzerland}
\author{B.~Deveaud-Pl\'edran} \affiliation{ICMP, Ecole Polytechnique F\'ed\'erale de Lausanne (EPFL), 1015 Lausanne, Switzerland}

\vskip 0.25cm

\date{\today}

\begin{abstract}

Exciton-polaritons have been shown to be an optimal system in order to investigate the properties of bosonic quantum fluids. We report here on the observation of dark solitons in the wake of engineered circular obstacles and their decay into streets of quantized vortices. Our experiments provide a time-resolved access to the polariton phase and density, which allows for a quantitative study of instabilities of freely evolving polaritons. The decay of solitons is quantified and identified as an effect of disorder-induced transverse perturbations in the dissipative polariton gas.

\end{abstract}

\maketitle

Perturbations of quantum fluids can lead to the creation of solitary waves called solitons resulting from the compensation between dispersion and particle interaction. In the particular case of repulsive interaction, dark solitons are created. These density depressions move in the fluid while keeping a constant shape and they are characterized by a phase jump across the density minimum. Since the first theoretical prediction \cite{tsuzuki_nonlinear_1971}, dark solitons have been studied and then observed in a variety of systems such as nonlinear lattices \cite{denardo_observations_1992}, thin magnetic films \cite{chen_microwave_1993} and complex plasma \cite{heidemann_dissipative_2009}. They have attracted considerable interest especially in the field of nonlinear optics \cite{kivshar_dark_1993}, because of their consequent use in communication devices (i.e optical fibers \cite{Mollenauer_1980}), and in atomic BEC \cite{frantzeskakis_dark_2010}.
As quantized vortices \cite{frisch_transition_1992,raman_vortex_2001,raman_evidence_1999}, dark solitons are BEC excitations, which arise spontaneously upon the phase transition. As such, they are clear evidences for the onset of a quantum behavior and powerful tools to understand BEC instabilities. Controlling these instabilities is of crucial importance for the development of optoelectronic devices based on quantum fluids in which stable regimes and structures are required.
Dark solitons are considered as the dispersive and nonlinear analog of shock waves of supersonic motion \cite{el_oblique_2006}. The creation of solitons by phase imprinting in BEC has been reported \cite{burger_dark_1999,denschlag_generating_2000}, triggering a growing interest in their hydrodynamic formation and stability of solitons. 

\begin{figure*}
\includegraphics[width=1\textwidth]{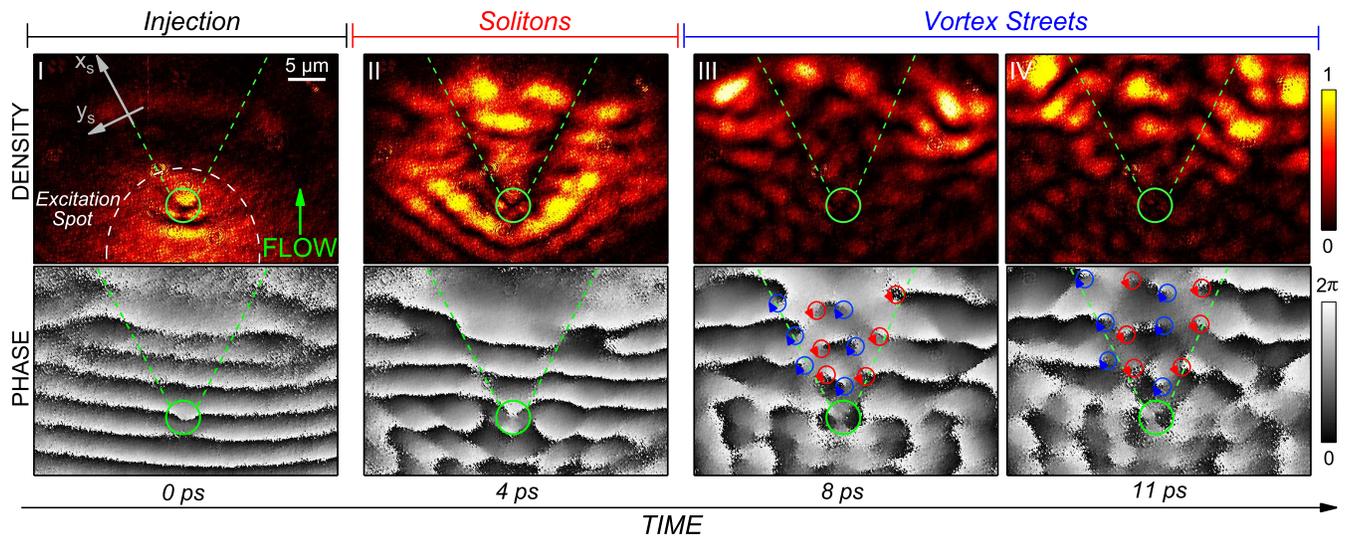}
\caption{Time evolution of density (upper panel) and phase (lower panel) of a polariton wavepacket scattering on an engineered circular obstacle (green circles) of 3 $\mu m$ diameter. Polaritons are injected at $t=0$ $ps$  with a 1.5 $\mu m ^{-1}$ initial momentum and a pump power of 5 $mW$ in the vicinity of the mesa (column I). Solitons (II) are visible few picoseconds later and are characterized by low density straight lines and a phase shift. The soliton decay (III) comes along with the breaking of the phase-fronts and the formation of vortex streets (blue circles). The latter is pointed out by the apparition of several density minima coinciding with phase singularities. The motion of vortices is tracked along the flow during polariton lifetime (IV).}
\end{figure*}

In this paper we report on the observation of hydrodynamic oblique dark solitons in a 2D polariton fluid and the formation of quantized vortex streets. 
Polaritons are bosonic quasi-particles arising from the strong coupling between quantum well excitons and photons in semiconductor microcavities. Due to their small effective mass and to their strong nonlinearity, they have turned out to be an optimal system in order to investigate the properties of a bosonic quantum fluid. Polaritons can undergo BEC \cite{kasprzak_bose-einstein_2006} and, by virtue of their non-equilibrium nature, they are accompanied with the spontaneous appearance of quantized vortices \cite{Lagoudakis2011}. With the demonstration of polariton superfluidity \cite{amo_superfluidity_2009,amo_collective_2009} much effort has been performed to better understand the nature of different turbulences arising from the breakdown of this fascinating state of matter. Recently, the hydrodynamic properties of a polariton superfluid have been investigated both theoretically \cite{pigeon_hydrodynamic_2010,Cuevas2011} and experimentally. Nucleation of quantized vortex pairs \cite{nardin_hydrodynamic_2011} and dark solitons \cite{amo_hydrodynamic_2011} have been observed in the wake of an obstacle inserted into the polariton fluid. Differently to the work reported in Ref.\cite{amo_hydrodynamic_2011}, in which a steady state was created by a continuous wave pumping, our pulsed experiments allow us to retrieve the dynamics of the polariton fluid and thus to reconstruct the temporal behavior of dark solitons. Time and phase resolution consents to observe the formation of vortex streets arising from the unstable nature of solitons and to quantitatively access the stability conditions.\newline
It is well known that dark solitons are unstable with respect to transverse perturbations \cite{kadomtsev_stability_1970,zakharov_instability_1975} and that they eventually decay into other more stable structures \cite{mcdonald_dark_1993,anderson_watching_2001,mironov_structure_2010}. The driven dissipative nature of polaritons results in the exploration of different hydrodynamic regimes enriching the achievable features of polariton-solitons. This allows in particular the study of soliton instability.\newline 

In our experiments we observe solitons in the wake of an engineered potential. The problem of a quantum fluid passing an obstacle can be formulated using the Gross-Pitaevskii equation and introducing an external potential term \cite{frisch_transition_1992}. In the case of a polariton gas, whose nature is strongly dissipative due to the finite lifetime, it has recently been proved that the ratio between the local flow velocity and local value of the speed of sound, namely the Mach number $M$, determines the hydrodynamic regimes \cite{pigeon_hydrodynamic_2010}. Solitons are predicted to appear for $M$ larger than 1 with values depending on the nature of the obstacle.  We resonantly inject polaritons in a GaAs microcavity with a single InGaAs quantum well featuring a 3.5 meV Rabi splitting \cite{kaitouni_engineering_2006}. The initial in-plane momentum is imposed by the pump. A 2 ps long pulse is used to create a polariton wavepacket which can subsequently evolve freely in the microcavity plane during the polariton lifetime and then show features typical of quantum fluid hydrodynamics \cite{nardin_hydrodynamic_2011}. In the microcavity a set of different engineered circular traps, called mesas, have been realized by etching the spacer before growing the top DBR \cite{kaitouni_engineering_2006}. They result in negative potentials barriers for the resonantly injected polariton wavepacket. Phase and amplitude of polaritons are resolved in time through homodyne detection. This is achieved by analyzing the spatial interference between a local oscillator and the polariton emission out of a Mach-Zehnder interferometer \cite{nardin_hydrodynamic_2011,nardin_phase-resolved_2010}. The dynamics is reconstructed from the subsequent interference images taken by varying the delay between the two arms of the interferometer. \newline

Figure 1 shows an example of the scattering dynamics of a supersonic polariton wavepacket against a 3 $\mu m$ mesa (green circles). Polaritons are injected with a 1.5 $\mu m ^{-1}$ initial momentum and a pump power of 5 $mW$ distributed over a laser spot of about 20 $\mu m$ diameter (in all panels flow is from the bottom to the top). These parameters unambiguously place the experiment in the supersonic regime \cite{nardin_hydrodynamic_2011}. The FWHM and the position of the laser spot with respect to the mesa is indicated by a white dashed line in the density profile of Fig. 1-I.  It can be observed in Fig.1-II that, due to the attractive nature of the obstacle, the injection pulse also excites confined polariton states close to the continuum \cite{nardin_phase-resolved_2010,cerna_coherent_2009}, which are characterized by bright lobes. \newline
Hydrodynamic soliton formation is the result of the interaction between the polariton fluid and the mesa, which provides a continuous perturbation on the condensate in motion.
The dynamics of a 2D fluid passing an obstacle and soliton formation can be analyzed in both lab and fluid reference frame. In the latter, the mesa moves with respect to the fluid leaving behind perturbations which expand as circular waves. Positive interference between them occurs tangentially resulting in the oblique soliton formation. They then move with respect to the fluid frame in the $y_s$ direction whereas they grow in $x_s$ direction in the lab frame (see Fig1-I).\newline
Such solitons are emphasized in Fig. 1-II by green dashed lines. Phase jumps across the amplitude minimum can be observed in the phase map.\newline
Few picoseconds after the polariton injection, solitons decay into vortex streets. The breaking of the phase-fronts creates singularities associated with density minima: a clear evidence for quantized vortices. These are highlighted in the phase-map by blue and red circles. Column III shows the density profile and the phase map during the formation of vortex-antivortex pairs along the streets. The motion of vortices can be tracked in order to retrieve their dynamics (Fig. 1-III and 1-IV).\newline 

\begin{figure}[t]
\includegraphics[width=1\columnwidth]{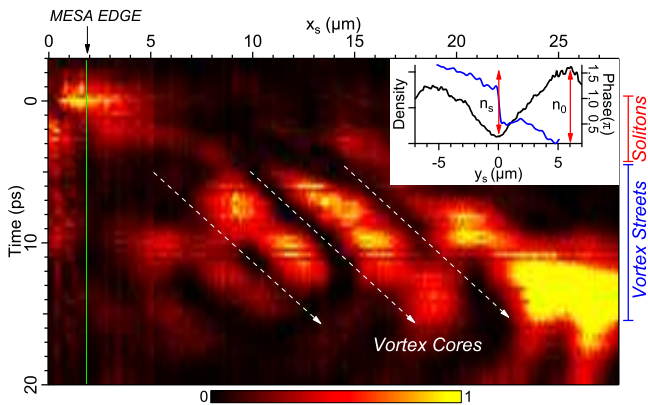}
\caption{Density map along the soliton direction $x_s$ versus time. After the pulse arrival at $t=0$ $ps$, the presence of a low density region stresses the nucleation of a soliton. Around $t=5$ $ps$ soliton decay occurs and the formation of vortices is highlighted by the appearance of periodic modulation of the density in the region previously occupied by the soliton. Insert plot shows the density (black) and the phase (blue) profile along $y_s$ at $t=3$ $ps$ calculated at around 7 $\mu m$ from the mesa center showing the characteristic soliton transversal shape and the phase jump across the minimum.}
\end{figure}

A unique way to investigate the soliton decay into vortex streets is given by the observation of the density changes in time along the soliton direction ($x_s$ in Fig. 1-I). A plot versus time of the density map along $x_s$ is shown in Figure 2. After the pulse arrival at $t=0$ $ps$, solitons nucleate until $t=4$ $ps$ and their presence is revealed by a long low density region. The insert of Figure 2 shows the characteristic soliton profile (black line) along its transversal direction $y_s$ at $t=3$ $ps$, before the formation of vortices. During the whole soliton lifetime, the soliton width has an almost constant value around 4.5 $\mu m$, not limited by the healing length and independent of the laser pump power. The phase jump along $y_s$ (namely across the soliton lines) typical of dark solitons is shown in the insert (blue line).\newline 
Around five picoseconds after the nucleation, solitons decay into vortex streets which are visualized by the appearance of periodic modulations of the density. The presence of vortices in Fig.2 is revealed by sloping stripes of alternate density. Low density valleys represent the motion of the cores of quantized votices. The size of the latter is estimated to be around 2 $\mu m$. Once formed, vortices move with a constant velocity $v=0.85$ $\mu m/ps $ along the $x_s$ axis, as demonstrated by the slope of the core trajectories in Fig.2. This value must be compared with the projection of flow velocity over $x_s$. The latter is calculated considering the ballistic propagation of polaritons with high momentum of injection \cite{Cerna_2010} and it matches with the one extracted from the phase map at the time of injection. The aperture angle of an oblique soliton with respect to the flow direction is measured to be $\sim 30^{\circ}$ at the decay time giving a component of $v_{flow}$ along $x_s$ of $0.95$ $\mu m / ps$. As expected, vortices move along the flow direction with a velocity comparable to that of the fluid.

The same experiment is repeated for a set of excitation powers ranging from 0.5mW to 5mW. Contrarily to what could come out from a na\"ive picture, we observe (see Fig. $3a$) that the solitons lines become less stable when the excitation power is increased (snapshots of the dynamics lower pump powers can be seen in the supplementary materials). The black dots in the Fig. $3a$ show the average time delay for the vortex street formation. The horizontal bars represent the time window during which the soliton instability is observed, namely the time between the formation of the first and the last of the vortex pairs. Note that we consider the whole instability with vortices appearing progressively all along the solitons in region where the density and thus the fluid behavior could be at different stage of the soliton evolution.\newline
Our observations are in agreement with previous studies on hydrodynamic nucleation of solitons and can be understood in terms of values of the Mach number in the obstacle perimeter. Solitons become more stable for higher values of $M$ \cite{el_oblique_2006,Kamchatnov_2008}. These are obtained experimentally by decreasing the sound speed, namely decreasing the polariton fluid density.

\begin{figure}[b]
\includegraphics[width=1\columnwidth]{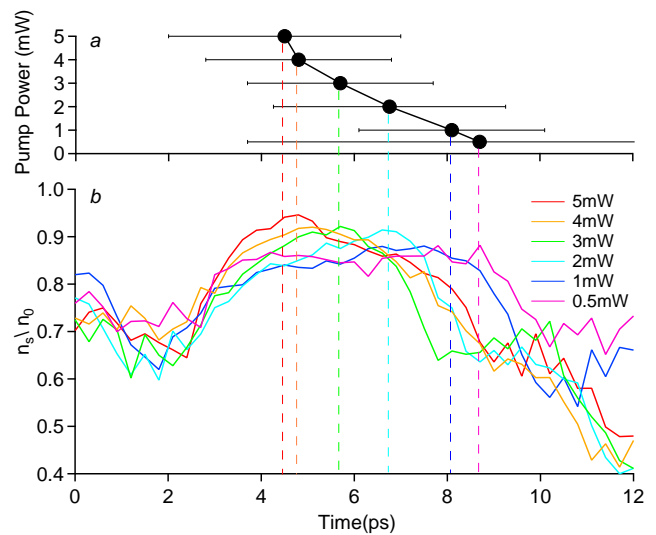}
\caption{{\bf a} - Pump power versus soliton lifetime. Horizontal bars show the time window at which vortex street formation occurs. Central black spots highlight that soliton lifetime increases while decreasing the pump power. {\bf b} - Time evolution of the soliton amplitude $n_s$ over polariton fluid density $n_0$ for different pump powers.  Maximum of the ratio is found in correspondence of the soliton decay into vortex streets (dashed vertical lines).}
\end{figure}

Studies on 1D solitons in atomic condensates revealed that the soliton speed $v_s$ is related to the ratio between the soliton amplitude and the fluid density $n_s/n_0$  through the relation $v_s/c_s=\sqrt{1-n_s/n_0}$ ($n_0$ and $n_s$ are measured as illustrated in the insert of Fig.2 and $c_s$ is the speed of sound) \cite{denschlag_generating_2000}. This means that $v_s$ basically vanishes when the density at the bottom of the soliton line becomes very small.

The time evolution of $n_s/n_0$  is shown in Fig.3b. It has been calculated by integrating the polariton emission over a 4 $\mu m$ long area at around 7 $\mu m$ away from the mesa. The Mach number is estimated from the values of $n_s/n_0$ to be larger than 2, which confirms that our experiments are placed in the supersonic regime.
During the first picoseconds of the evolution, $n_s/n_0$ increases for all pump powers, showing that the solitons are becoming deeper. Remarkably, the observed break-up of the solitons corresponds quite precisely to the moment when this ratio reaches a critical value of 0.9 for all experimental conditions. In disorder-free media, soliton instability is predicted to appear when $n_s/n_0$ reaches 1. It has also been shown that potential fluctuations modulate the fluid density and thus the local soliton velocity \cite{kuznetsov_instability_1988,kuznetsov_instability_1995}. Finally, in the case of soliton-like bubbles \cite{Barashenkov_1988}, the critical ratio for stability is found to be smaller than 1 , depending on the surrounding potential. From these theoretical inputs, we conclude that the critical ratio of 0.9 in our experiments can be attributed to the photonic disorder, typical of semiconductor microcavities.

Our observations are supported by theoretical investigations performed by solving iteratively the generalized dissipative Gross-Pitaevskii equation for the lower polariton branch \cite{Carusotto2010}. Details about the methods can be found in Ref.\cite{nardin_hydrodynamic_2011}. Solitons nucleation in the wake of a negative potential and the subsequent decay into vortex streets is very well reproduced by the numerical simulations which take into account the pulsed excitation, the finite spot size of the laser, the exponential decay of the polariton population and the disorder potential. Snapshots of the simulated dynamics are reported in Fig.4. They reproduce perfectly the timescale of the experiments. After the polariton injection at $t=0$ $ps$, solitons are created almost instantly and transform in vortex streets after about $t=5$ $ps$. After creation, vortices move according to the flow as observed in the experiments. It is interesting to notice that, while keeping constant all the other parameters , no soliton decay is observed when the disorder potential amplitude is lowered to zero. 

\begin{figure}[t]
\includegraphics[width=1\columnwidth]{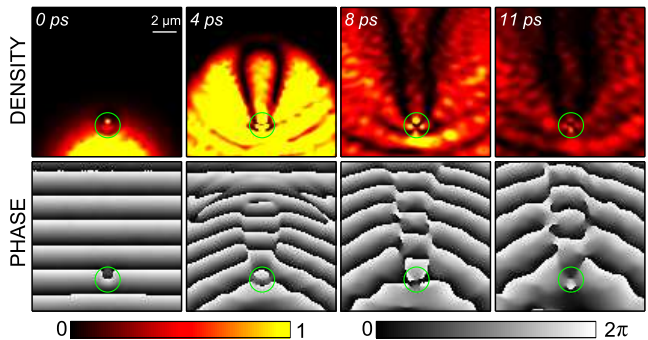}
\caption{Snapshots of the numerical simulations, based on the generalized Gross-Pitaevskii equations for the nucleation and the decay of dark solitons. The 2 $\mu m$ diameter negative potential of $-4$ meV is indicated by a green cirlce. Flow direction is upstream. Random potential disorder is added in order to perturb solitons.}
\end{figure}

Other similar experiments have been performed for mesas of different dimensions with varying initial flow velocity (see Fig.3S in online supplementary materials). In agreement with a previous work \cite{nardin_hydrodynamic_2011}, vortex pairs are nucleated for smaller injection momentum. On the other hand, an irregular regime with the nucleation of many quantized vortices is observed for a mesa with a diameter of 9 $\mu m$ (see Fig.4S in online supplementary materials), as predicted for atomic BEC \cite{sasaki_benardvon_2010}.

In conclusion, we have observed, in a dissipative polariton gas, oblique dark solitons resulting from the scattering of a polariton wavepacket against an engineered circular attractive obstacle. Soliton properties can be controlled by tuning the initial density and velocity of the fluid. Stability conditions has been quantitatively investigated due to the time resolved polariton phase and amplitude. Solitons tend to reach the maximum possible value of the ratio $n_s/n_0$ within a time that gets shorter for higher excitation density. At such a point, cavity disorder was found to be responsible for the soliton instability and their decay into vortex streets.

We would like to thank O. El Da\"if for the patterning of the microcavity sample and N. Berloff and C. Ciuti for fruitful discussions about quantum hydrodynamics. We acknowledge support by the Swiss National Science Foundation through the "NCCR
Quantum Photonics."


\end{document}